\documentstyle[twocolumn,prd,aps]{revtex}

\newcommand{\K}{{\cal K}}

\begin{document} \draft


\wideabs{
\title{Kasner-AdS spacetime and anisotropic brane-world cosmology}
\author{Andrei V. Frolov}
\address{
  CITA, University of Toronto\\
  Toronto, Ontario, Canada, M5S 3H8\\
  {\rm E-mail: \texttt{frolov@cita.utoronto.ca}}
}
\date{13 February 2001}
\maketitle

\begin{abstract}
  Anisotropic generalization of Randall and Sundrum brane-world model
  is considered. A new class of exact solutions for brane and bulk
  geometry is found; it is related to anisotropic Kasner solution. In
  view of this, the old question of isotropy of initial conditions in
  cosmology rises once again in the brane-world context.
\end{abstract}

\pacs{PACS numbers: 04.50.+h, 98.80.Cq \hfill CITA-2001-05}
}
\narrowtext


\section{Introduction}

String theory suggests that the spacetime we live in might be
fundamentally higher-dimensional \cite{Horava:96a,Horava:96b}. Some of
these extra dimensions might be compactified to account for apparently
four-dimensional spacetime we observe experimentally. Recently, Randall
and Sundrum proposed a new model with relatively large extra dimension
\cite{Randall:99a,Randall:99b} as a way to solve the hierarchy problem
in high energy physics. In this model, the matter fields and
interactions with exception of gravity are localized on 3-branes that
live in a 5-dimensional bulk spacetime, which is taken to be anti-de
Sitter (AdS).

The Randall and Sundrum model has gained considerable popularity, in
both high energy physics and cosmology communities. A number of
cosmological scenarios based on the brane-world concept was explored
\cite{Binetruy:99a,Binetruy:99b,Shiromizu:99,Mukohyama:99a,Mukohyama:99b},
including some models with inflation on the brane
\cite{Dvali:98,Maartens:99,Copeland:00,Himemoto:00} and different
embedding geometries \cite{Maartens:00,Campos:01}.

In the present letter, we explore possible anisotropic brane-world
cosmologies. We obtain a new class of exact solutions to the
5-dimensional vacuum Einstein-AdS equations in the bulk, which is
homogeneous but anisotropic, and is related to Kasner solution. We then
consider brane embedding in this anisotropic bulk spacetime, and derive
brane equations of motion. These are solved by a static brane
configuration with the brane tension tuned to the Randall-Sundrum
prescription. The geometry on the 3-brane is given by 4-dimensional
Kasner solution.

We also discuss implications of existence of these anisotropic
solutions to the brane-world cosmology, and ponder possible ways to
solve the initial conditions problem \cite{Lifshitz:63,LL:FT} in the
brane-world context.

\section{Kasner-AdS spacetime}

In the usual brane world scenario \cite{Randall:99a,Randall:99b}, the
$3$-branes live in a $5$-dimensional AdS bulk spacetime, which is
described by a metric
\begin{equation} \label{eq:metric}
  ds^2 = -f(r) dt^2 + \frac{dr^2}{f(r)} + r^2 d\sigma_3^2,
\end{equation}
where $d\sigma_3^2$ is a $3$-dimensional metric of a unit sphere, plane
or hyperboloid for $K=+1, 0, -1$ respectively, and
\begin{equation} \label{eq:f}
  f(r) = K + \frac{r^2}{\ell^2},
\end{equation}
with $\ell$ giving the curvature scale of the AdS spacetime. This bulk
spacetime is a solution of vacuum Einstein-AdS equations
\begin{equation} \label{eq:einstein}
  R_{\mu\nu} - \frac{1}{2} R g_{\mu\nu} =  -\Lambda g_{\mu\nu},
\end{equation}
with negative cosmological constant
\begin{equation} \label{eq:lambda}
  \Lambda = - \frac{6}{\ell^2}.
\end{equation}

We now generalize the spacetime (\ref{eq:metric}) to include spatial
anisotropy. For simplicity, we will only consider spatially flat case
($K=0$), for which the spatial part of AdS metric was
\begin{equation} \label{eq:metric:flat}
  d\sigma_3^2 = dx^2 + dy^2 + dz^2.
\end{equation}
In order to introduce anisotropy to the bulk spacetime, yet keep the
spatial slices homogeneous, we allow the coefficients of the spatial
metric to vary with time
\begin{equation} \label{eq:metric:anisotropic}
  d\sigma_3^2 = e^{2a(t)} dx^2 + e^{2b(t)} dy^2 + e^{2c(t)} dz^2.
\end{equation}
The vacuum Einstein-AdS equations (\ref{eq:einstein}) then give the
evolution of anisotropy scales $a$, $b$ and $c$, which obey three
dynamical equations of motion, namely
\begin{equation}
  \ddot{a} + \dot{a}^2 - \dot{b}\dot{c} = 0
\end{equation}
and permutations thereof with respect to interchanges of $\{a,b,c\}$,
and a constraint
\begin{equation}
  \dot{a}\dot{b} + \dot{b}\dot{c} + \dot{c}\dot{a} = 0,
\end{equation}
where dot denotes derivatives with respect to time $t$. The general
solution of these equations, modulos the translational and scaling
freedom in the coordinate choice, is
\begin{equation} \label{eq:soln}
  a = \alpha \ln t, \hspace{1em}
  b = \beta \ln t, \hspace{1em}
  c = \gamma \ln t,
\end{equation}
where parameters $\alpha$, $\beta$ and $\gamma$ satisfy either
\begin{equation}
  \alpha = \beta = \gamma = 0,
\end{equation}
in which case the original flat spatial metric (\ref{eq:metric:flat})
is recovered, or
\begin{equation} \label{eq:parameters}
  \alpha^2 + \beta^2 + \gamma^2 = \alpha + \beta + \gamma = 1.
\end{equation}
The spatial geometry in this case is homogeneous but anisotropic, and
is given by the metric
\begin{equation} \label{eq:metric:kasner}
  d\sigma_3^2 = t^{2\alpha} dx^2 + t^{2\beta} dy^2 + t^{2\gamma} dz^2.
\end{equation}

Thus we have obtained a new solution
(\ref{eq:metric:kasner},\ref{eq:parameters},\ref{eq:metric}) of five
dimensional vacuum Einstein-AdS equations (\ref{eq:einstein}). It is
related to the well-known Kasner solution of vacuum Einstein equations
in four dimensions (indeed, the three dimensional spatial line elements
are identical), so we will call it a Kasner-AdS spacetime. Its
properties with respect to spatial anisotropy are similar to those of
Kasner spacetime. However, its global structure is different in that it
not only has a cosmological singularity at $t=0$, but also a timelike
singularity at $r=0$, where the curvature diverges
\begin{equation}
  C_{\alpha\beta\gamma\delta} C^{\alpha\beta\gamma\delta} =
    - \frac{16\ell^4}{r^4 t^4}\, \alpha\beta\gamma.
\end{equation}
This does not pose a significant problem for the brane-world scenario,
however, as the central part of the spacetime is avoided in orbifold
construction \cite{Randall:99a}.

It is worth noting that similar generalization to 5-dimensional AdS
theory is possible for any Ricci-flat 4-dimensional metric,
Schwarzschild black hole in particular \cite{Birmingham:98}. It might
be interesting to look for spatially closed anisotropic solutions, as
they are likely to be of Mixmaster type and chaotic. However, this
topic is beyond the scope of this letter. For several other interesting
generalizations of Kasner solution, see
\cite{Kokarev:95,Dabrowski:99,Halpern:00}.

\section{Brane in anisotropic bulk}

We now consider what will happen if the $3$-brane is embedded in the
anisotropic Kasner-AdS spacetime derived above, instead of the usual
AdS spacetime. Following \cite{Mukohyama:99b}, we describe the
$3$-brane embedding by a hypersurface $\Sigma$ defined by $r=R(t)$.
Induced brane-world metric is then
\begin{eqnarray} \label{eq:metric:brane}
  ds_4^2
    &=& - \left(f - \frac{R_{,t}^2}{f}\right) dt^2 + R^2 d\sigma_3^2 \nonumber\\
    &=& - d\tau^2 + A^2 d\sigma_3^2,
\end{eqnarray}
where we introduced cosmological time
\begin{equation}
  d\tau = \left(f - \frac{R_{,t}^2}{f}\right)^{\frac{1}{2}} dt, \hspace{1em}
  dt = \frac{1}{f} \left(f + A_{,\tau}^2\right)^{\frac{1}{2}} d\tau
\end{equation}
and cosmological scale factor
\begin{equation}
  A(\tau) = R(t(\tau)).
\end{equation}
The coordinates on the brane are $\{\tau, x, y, z\}$, and the
corresponding holonomic basis vectors are
\begin{equation} \label{eq:basis}
  e_{(\tau)}^\mu = \left(\frac{1}{f} (f + \dot{A}^2)^{\frac{1}{2}}, \dot{A}, 0, 0, 0 \right)
\end{equation}
and $e_{(a)}^\mu = \delta_a^\mu$ for index $a$ spanning $\{x, y, z\}$.
Here and later the dot denotes the derivative with respect to
cosmological time $\tau$. The outward pointing unit vector normal to
the brane is
\begin{equation} \label{eq:normal}
  n^\mu = \left(\frac{\dot{A}}{f}, (f + \dot{A}^2)^{\frac{1}{2}}, 0, 0, 0 \right).
\end{equation}
By direct calculation, the extrinsic curvature tensor for the brane
embedded in the spacetime (\ref{eq:metric},\ref{eq:metric:anisotropic}),
defined as
\begin{equation} \label{eq:K}
  \K_{ab} = e_{(a)}^\mu e_{(b)}^\nu n_{\mu;\nu},
\end{equation}
has the following nonvanishing components:
\begin{equation} \label{eq:Ktt}
  \K_{\tau\tau} = - \frac{1}{2}\, \frac{\ddot{A} + f'}{\sqrt{f + \dot{A}^2}},
\end{equation}
\begin{equation} \label{eq:Kxx}
  \K_{xx} = Ae^{2a} \left( \frac{A\dot{A}}{f} \frac{\partial a}{\partial t} + \sqrt{f + \dot{A}^2} \right),
\end{equation}
as well as $\K_{yy}$ and $\K_{zz}$, which are the same as the expression
for $\K_{xx}$ above, except anisotropy scale $a$ is replaced by $b$ and
$c$ correspondingly.

Assuming $Z_2$ symmetry common to the brane-world models
\cite{Horava:96a,Horava:96b,Randall:99a,Randall:99b,%
Binetruy:99a,Binetruy:99b,Shiromizu:99,Mukohyama:99a,Mukohyama:99b},
the jump in extrinsic curvature across the brane is
\begin{equation}
  [\K_{ab}] = \pm 2 \K_{ab},
\end{equation}
where sign depends on which side of the brane the bulk is (plus if
bulk is outside, i.e. towards the larger $r$, and minus if bulk is
inside, i.e. towards smaller $r$).

The jump in extrinsic curvature is caused by the matter distribution on
the brane; more precisely, Israel's junction condition \cite{Israel:66}
links it with the surface stress-energy tensor
\begin{equation} \label{eq:junction}
  [\K_{ab}] = - \kappa_D^2 \left(S_{ab} - \frac{1}{D-2}\, S g_{ab}\right),
\end{equation}
where $D$ is the dimensionality of the spacetime ($D=5$ in our case),
and $\kappa_D^2$ is $D$-dimensional Newton's constant. Together with
expressions for extrinsic curvature (\ref{eq:Ktt},\ref{eq:Kxx}),
junction condition (\ref{eq:junction}) gives equations of motion of the
brane, provided the brane matter content is known.

Assuming the matter on the brane is composed of vacuum energy of
density $\lambda$ and pressureless matter of density $\rho$, the
stress-energy tensor and its trace are
\begin{equation} \label{eq:S}
  S^a_b = \text{diag}(-\rho-\lambda, -\lambda, -\lambda, -\lambda), \hspace{1em}
  S = -\rho-4\lambda.
\end{equation}
The brane equations of motion are then
\begin{equation} \label{eq:eom:tt}
  \frac{\ddot{A} + f'}{\sqrt{f + \dot{A}^2}} = \pm \frac{\kappa_5^2}{3} (2\rho-\lambda),
\end{equation}
\begin{equation} \label{eq:eom:xx}
  \frac{\dot{A}}{f} \frac{\partial a}{\partial t} + \sqrt{\frac{f}{A^2} + \frac{\dot{A}^2}{A^2}} =
    \mp \frac{\kappa_5^2}{6} (\rho+\lambda),
\end{equation}
plus two equations coming from $yy$ and $zz$ components of $[\K_{ab}]$,
which are identical to the last equation, except anisotropy scale $a$
is replaced by $b$ and $c$ correspondingly. Since for anisotropic bulk
$a$, $b$ and $c$ depend on time differently (as given by equation
(\ref{eq:soln})), the only way to satisfy all equations simultaneously
without introducing anisotropic matter content on the brane, is to have
the anisotropic term in (\ref{eq:eom:xx}) vanish. This is achieved when
$A=\text{const}$, i.e. when brane is not moving. In that case, the
junction conditions (\ref{eq:eom:tt},\ref{eq:eom:xx}) simplify greatly
and become
\begin{equation}
  \frac{1}{\ell} =
    \pm \frac{\kappa_5^2}{6} (2\rho-\lambda) = 
    \mp \frac{\kappa_5^2}{6} (\rho+\lambda).
\end{equation}
They are satisfied (and thus anisotropic brane-world construction is
possible) when the brane parameters are
\begin{equation}
  \lambda = \mp \frac{6}{\kappa_5^2 \ell}, \hspace{1em}
  \rho = 0,
\end{equation}
which is precisely the tuning condition of Randall and Sundrum 
\cite{Randall:99a,Randall:99b}.

\section{Discussion}

We have shown that the geometrical construction of the Randall and
Sundrum's brane world \cite{Randall:99a} can be carried out without
assumption of spatial isotropy. This results in a wider class of
solutions of Einstein-AdS equations in five dimensions that are
anisotropic in the bulk and have Kasner geometry on the brane.
Written in terms of fifth coordinate $w = -\ell \ln(r/\ell)$ that
Randall and Sundrum use, the spacetime metric is
\begin{equation}
  ds^2 = e^{-2w/\ell} ( -dt^2 + t^{2\alpha} dx^2 + t^{2\beta} dy^2 + t^{2\gamma} dz^2) + dw^2.
\end{equation}

In view of existence of such solutions, the old problem of initial
conditions in cosmology \cite{Lifshitz:63,LL:FT} rises once again, now
in the brane-world context. The standard AdS brane-world solution is
highly symmetrical and essentially forms a set of measure zero among
the configuration space of possible solutions. Why should we be living
in such a special spacetime? This requires justification: for example,
a mechanism by which homogeneous and isotropic AdS brane-world can be
reached from wide range of initial conditions.

In the accepted 4-dimensional cosmology, the answer to the problem of
initial conditions was provided by inflation. Inflation on the brane
has been considered in the literature (see for example
\cite{Dvali:98,Maartens:99,Copeland:00,Himemoto:00}), and even the
tendency of the brane embedded into AdS bulk to dissipate anisotropy
was reported \cite{Maartens:00,Campos:01}, but it is not perfectly
clear what would happen if \textit{the bulk itself} was anisotropic (or
non-homogeneous, for that matter). Would inflation on the brane be
enough to isotropize both the brane \textit{and} the bulk (which has
much larger volume), or would some sort of ``bulk inflation'' be
needed?

These questions require further study.

\section*{Acknowledgments}

This research was supported in part by the Natural Sciences and
Engineering Research Council of Canada. I would like to thank Prof. Lev
Kofman for many helpful discussions.



\end{document}